\documentclass[twoside]{LCWS11}
\usepackage[latin1]{inputenc}
\usepackage[dvips]{graphicx,epsfig,color}
\usepackage{wrapfig,rotating}
\usepackage{amssymb,amsmath,array}
\usepackage{cite}

\usepackage{subfigure}
\usepackage{rotating,multirow,booktabs}

\newcommand{\mneu}[1]{m_{\tilde{\chi}^0_{#1}}}
\newcommand{\neu}[1]{\tilde{\chi}^0_{#1}}
\newcommand{\fb}{\ \mathrm{fb}}
\newcommand{\gev}{\ \mathrm{GeV}}

\begin{document}
\title{Measurement of CP Violation in the MSSM Neutralino Sector with
  the ILD}
%***********************************************************************
% AUTHORS INFORMATION AREA
%***********************************************************************
\author{M.~Terwort$^1$, O.~Kittel$^2$, G.~Moortgat-Pick$^{1,3}$, K.~Rolbiecki$^1$ and P.~Schade$^{1,4}$
\vspace{.3cm}\\
1- DESY, Notkestra{\ss}e 85, D-22607 Hamburg, Germany
\vspace{.1cm}\\
2- Departamento de F\'isica Te\'orica y del Cosmos and CAFPE,\\
\vspace{.1cm}
Universidad de Granada, E-18071 Granada, Spain\\
\vspace{.1cm}
3- University of Hamburg, Luruper Chaussee 149, D-22761 Hamburg, Germany\\
\vspace{.1cm}
4- CERN, CH-1211 Geneve 23, Switzerland\\
}
%%***********************************************************************
% END OF AUTHORS INFORMATION AREA
%***********************************************************************

\maketitle

\begin{abstract}

  Supersymmetric models provide many new complex phases which lead to
  CP violating effects in collider experiments. As an example,
  CP-sensitive triple product asymmetries in neutralino production
  $e^+\,e^- \to\tilde{\chi}^0_i \, \tilde{\chi}^0_1$ and subsequent
  leptonic two-body decays $\tilde\chi^0_i \to \tilde\ell_R \, \ell$,
  $ \tilde\ell_R \to \tilde\chi^0_1 \, \ell$, for $ \ell= e,\mu$, are
  studied within the Minimal Supersymmetric Standard Model. A full ILD
  detector simulation has been performed at a center of mass energy of
  $\sqrt{s}=500$~GeV, including the relevant Standard Model background
  processes, a realistic beam energy spectrum, beam backgrounds and a
  beam polarization of 80\% and $-60\%$ for the electron and positron
  beams, respectively. Assuming an integrated luminosity of $500~{\rm
    fb}^{-1}$ collected by the experiment and the performance of the
  current ILD detector, a relative measurement accuracy of 10\% for
  the CP-sensitive asymmetry can be achieved in the chosen scenario.

\end{abstract}

\section{Introduction}

\begin{wrapfigure}{r}{0.4\columnwidth}
\centering
\includegraphics[width=0.35\columnwidth]{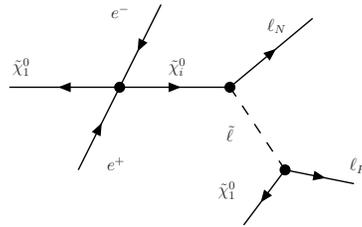}
\caption{\label{shematic picture}
Schematic picture of neutralino production and decay.}
\end{wrapfigure}

Supersymmetry (SUSY)~\cite{SUSY} is among the most favoured and most
studied extensions of the Standard Model (SM) and is capable of
solving many of its problems. One of its features is that the Minimal
Supersymmetric Standard Model (MSSM) provides a number of complex
parameters which can serve as sources of CP violation. They are
conventionally chosen to be the Higgsino mass parameter, $\mu = |\mu|
e^{i \phi_\mu}$, the ${\rm U(1)}$ and ${\rm SU(3)}$ gaugino mass
parameters, $M_1 = |M_1| e^{i \phi_{1}}$ and $M_3 = |M_3| e^{i
  \phi_3}$, respectively, and the trilinear scalar coupling
parameters, $A_f = |A_f| e^{i \phi_{A_f}}$, of the third generation
sfermions ($f=b,t,\tau$). CP phases can give rise to CP-violating
signals in collider experiments~\cite{Sabine}, which have to be
measured to determine or constrain the phases independently of
measurements of electric dipole moments (EDM). Although also CP-even
observables, such as masses or branching ratios, are sensitive to the
CP phases, CP-odd observables are needed for direct evidence of CP
violation.

In this report neutralino pair production $e^+\,e^-
\to\tilde{\chi}^0_i \, \tilde{\chi}^0_1$, for $i=2,3$, and the
subsequent leptonic two-body decay of one of the neutralinos
$\tilde\chi^0_i \to \tilde\ell_R \, \ell$ followed by $ \tilde\ell_R
\to \tilde\chi^0_1 \, \ell$, for $ \ell= e,\mu$, at the ILC is
studied~\cite{CP_Paper}. Figure~\ref{shematic picture} shows a
schematic picture of the process. The CP-sensitive spin correlations
of the neutralino in its production process allow to probe the phase
of the Higgsino mass parameter $\mu$ and the gaugino parameter
$M_1$~\cite{Bartl}.

A full ILD~\cite{ILD} detector simulation is performed in order to
investigate in detail the prospects to measure CP-sensitive
observables at the ILC. All relevant SM background is taken into
account, simulated with a realistic beam energy spectrum and beam
backgrounds~\cite{Peter}.

\section{CP-odd observables and benchmark scenario}\label{Sec:Scenario}

In neutralino production, effects from CP-violating phases can only
occur if two different neutralinos are produced. CP asymmetries can
then be defined with triple products of particle momenta. Due to the
spin correlation the asymmetries show hints for CP phases already at
tree level. For the process shown in Fig.~\ref{shematic picture}, a
T-odd triple product of the beam and the final lepton momenta can be
defined as~\cite{CP_Paper}
 \begin{eqnarray}
{\mathcal T} &=& 
        ({\mathbf p}_{e^-} \times {\mathbf p}_{\ell^+}) \cdot {\mathbf p}_{\ell^-}.
	\label{AT}
\end{eqnarray}
The corresponding asymmetry is
\begin{eqnarray}
  {\mathcal A}({\mathcal T} )  &=&  
        \frac{\sigma({\mathcal T}>0) - \sigma({\mathcal T}<0)}
             {\sigma({\mathcal T}>0) + \sigma({\mathcal T}<0)},
\label{eq:asyth}
\end{eqnarray}
where $\sigma$ is the cross section for neutralino production and
decay. Its sign depends on the charge of the leptons, which has to be
tagged in the experimental analysis.

For the full simulation study a benchmark scenario has been chosen
such that the gaugino phase $\phi_1=0.2\pi$ corresponds to a maximal
CP asymmetry and the Higgsino phase is zero, since it is strongly
constrained by EDM bounds. The other parameters in the neutralino
sector are $M_2=300$\,GeV, $|M_1|=150$\,GeV, $|\mu|=165$\,GeV and
$\tan{\beta}=10$. This leads to the neutralino masses
$m_{\tilde{\chi}^0_1} = 117~{\rm GeV}$, $m_{\tilde{\chi}^0_2} =
169~{\rm GeV}$, $m_{\tilde{\chi}^0_3} = 181~{\rm GeV}$ and
$m_{\tilde{\chi}^0_4} = 330~{\rm GeV}$, while the slepton masses are
$m_{\tilde \ell_R} = 166~{\rm GeV}$ and $m_{\tilde \ell_L} = 280~{\rm
  GeV}$. The neutralino pair production cross sections are calculated
to be $\sigma(e^+e^-\to\tilde\chi_1^0\tilde\chi_2^0) = 244~{\rm fb}$
and $\sigma(e^+e^-\to\tilde\chi_1^0\tilde\chi_3^0) = 243~{\rm fb}$,
while the slepton pair production cross sections are
$\sigma(e^+e^-\to\tilde{e}^+_R\tilde{e}^-_R) = 304~{\rm fb}$ and
$\sigma(e^+e^-\to\tilde{\mu}^+_R\tilde{\mu}^-_R) = 97~{\rm fb}$.  The
slepton pair production is the main background, since there are two
lightest neutralinos and two opposite-sign electrons or muons in the
final state as in the case of the neutralino
$\tilde\chi_1^0\tilde\chi_i^0$ production. Furthermore, beam
polarizations of $(P_{e^-},P_{e^+})=(0.8,-0.6)$ have been chosen,
which enhance slightly the SUSY cross section and the asymmetries,
while the background from $WW$- and chargino-pair production is
suppressed. In this scenario the CP asymmetries are ${\mathcal
  A}({\mathcal T} )_{\tilde\chi_1^0\tilde\chi_2^0} = -9.2\%$ and
${\mathcal A}({\mathcal T} )_{\tilde\chi_1^0\tilde\chi_3^0} = 7.7\%$.

\section{Detector simulation study and parameter fit}

The ILD is a concept under study for a multipurpose particle detector
for the ILC. It is designed for an excellent precision in momentum and
energy measurement over a large solid angle. A detailed description
can be found in~\cite{ILD}. In the simulation all active elements and
also cables, cooling systems, support structures and dead regions are
taken into account~\cite{Peter}. The radiation hard beam calorimeter
is used to suppress background from $\gamma\gamma$ events at low
angles. All relevant SM backgrounds and SUSY processes are generated
using \texttt{Whizard}~\cite{Whizard}.

\subsection{Event selection and measured asymmetry}

A clean sample of signal events is needed in order to clearly measure
the CP-violating effects in neutralino production. Otherwise the
asymmetry will be reduced by the CP-even background events. Therefore,
preselection cuts as listed in Tab.~\ref{tab:preselectionCuts} are
applied to reject as much background as possible, while preserving
good signal efficiency. Electrons and muons are identified using the
{\it Particle Flow} approach~\cite{CP_Paper}. The cuts exploit the
energy and angular distributions of the final state leptons, as well
as the high missing transverse momentum ${\mathbf p}_{\rm T}^{\rm
  miss}$ due to the escaping neutralinos. Additional cuts on the total
visible energy $E_{\rm vis}$ as well as on the invariant mass
$m_{\ell\ell}$ distributions further reduce the background
contamination.

\begin{table}%[htbp]
\begin{center}
%\vspace{0.5cm}
\renewcommand{\arraystretch}{1.3}
  \begin{tabular}{ll} \toprule
 initial selection & no significant activity in BCAL \\
                   & number of all tracks $N_{\rm tracks} \leq 7$ \\ \hline
 lepton selection  & $\ell^+\ell^-$ pair with $\ell = e,\mu$     \\
                   & $|\cos\theta|<0.99$, min. energy $E>3$~GeV \\
                   &  lower energetic $\ell$ with $E<18$~GeV, or \\ 
                   & \phantom{x} higher energetic $\ell$ with $E>38$~GeV \\
                   & higher energetic $\ell$ with $E\in[15,150]$~GeV \\
                   & $\theta_{\rm acop}>0.2\pi$, $\theta_{\rm acol}>0.2\pi$ \\ \hline
final preselection & ${\mathbf p}_{\rm T}^{\rm miss}> 20$~GeV  \\
                   & $E_{\rm vis}< 150$~GeV    \\
                   & $m_{\ell\ell}<55$~GeV \\ \bottomrule
  \end{tabular}
\end{center}
\renewcommand{\arraystretch}{1.0}
\caption{Preselection cuts, see Ref.~\cite{CP_Paper} for details.\label{tab:preselectionCuts}}
\end{table}

\begin{figure}[t]
  \centering
  \subfigure[]{\label{fig:p_T_miss_distribution}
    \includegraphics[width=2.5in]{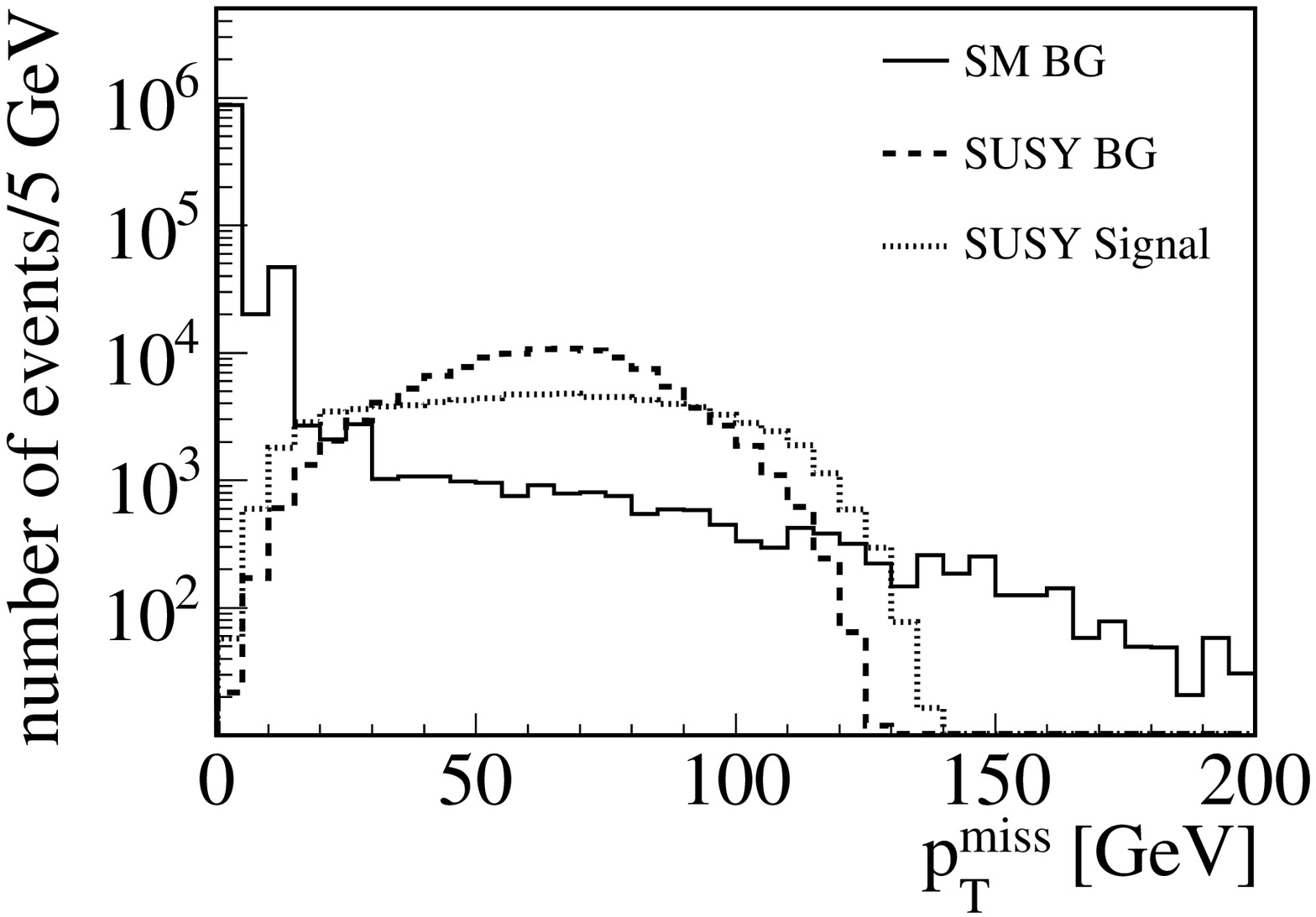}
  }
  \subfigure[]{\label{fig:InvMass_distribution}
    \includegraphics[width=2.5in]{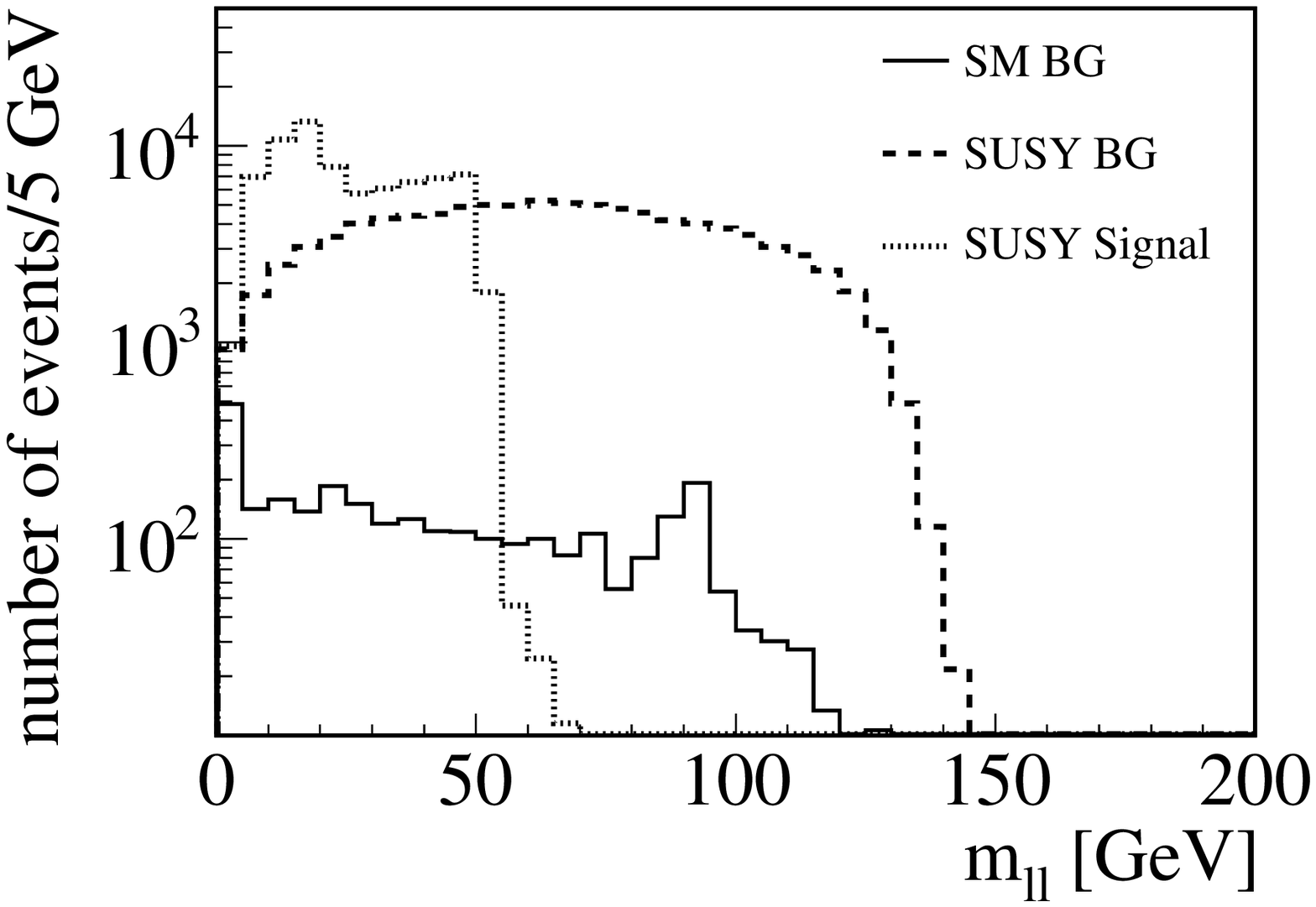}
  }
  \caption{ (a) Missing transverse momentum ${\mathbf p}_{\rm T}^{\rm
      miss}$ distribution of SM background, SUSY background and SUSY
    signal after the lepton selection. (b) Invariant mass
    $m_{\ell\ell}$ distribution of the lepton pair after all
    preselection cuts except the cut on $m_{\ell\ell}$. The events are
    simulated for ${\mathcal{L}} = 500 \fb^{-1}$, beam polarization
    $(P_{e^-}, P_{e^+}) = (0.8,-0.6)$ at $\sqrt{s}= 500$~GeV, and MSSM
    parameters as in the benchmark scenario discussed in
    Sec.~\ref{Sec:Scenario}.}
\label{fig:distributions}
\end{figure}

Figure~\ref{fig:p_T_miss_distribution} shows the ${\mathbf p}_{\rm
  T}^{\rm miss}$ distribution of the SM and SUSY background as well as
of the signal after the lepton selection. It can be seen that most of
the background is removed with the cut ${\mathbf p}_{\rm T}^{\rm
  miss}> 20$~GeV. Figure~\ref{fig:InvMass_distribution} shows the
distribution of the invariant di-lepton mass after all cuts except the
one on $m_{\ell\ell}$. The signal lepton pair from $\tilde{\chi}^0_3$
($\tilde{\chi}^0_2$) decays has a sharp endpoint at $51$~GeV
($22$~GeV), which can also be exploited for mass measurements. The
invariant mass cut also removes SM backgrounds from $ZZ$ and $WW$
production. The remaining event sample consists of 28039
$\tilde\chi^0_1\tilde\chi^0_2\rightarrow\tilde\chi^0_1\tilde\chi^0_1\ell\ell$
($\ell\neq\tau$) events, 45966
$\tilde\chi^0_1\tilde\chi^0_3\rightarrow\tilde\chi^0_1\tilde\chi^0_1\ell\ell$
($\ell\neq\tau$) events and 34223
$\tilde\ell\tilde\ell\rightarrow\tilde\chi^0_1\tilde\chi^0_1\ell\ell$
($\ell\neq\tau$) events. All other SM and SUSY background processes
sum up to about 6000 events.

In order to distinguish the $\tilde\chi^0_1\tilde\chi^0_2$ events from
the $\tilde\chi^0_1\tilde\chi^0_3$ events, and to further clean the
event sample, a kinematic selection procedure is applied, as described
in~\cite{CP_Paper}. A number of kinematic constraints derived from the
final state momenta are used to classify events as signal or
background. An event is selected only if it is classified exclusively
as signal-like. Four event classes are considered:
$\tilde\chi^0_1\tilde\chi^0_2$, $\tilde\chi^0_1\tilde\chi^0_3$,
$\tilde\ell^+_R\tilde\ell^-_R$ and $W^+W^-$.
Table~\ref{tab:reconstruction_results} shows the number of events that
are classified exclusively as one of the four event classes. It can be
observed that the large contamination of the event sample by
$\tilde\ell^+_R\tilde\ell^-_R$ events can be drastically reduced.

\begin{table}%[htbp]
\begin{center}
%\vspace{0.5cm}
\renewcommand{\arraystretch}{1.2}
  \begin{tabular}{lcccc} \toprule
    class  & only $\tilde\chi^0_1\tilde\chi^0_2$ & only $\tilde\chi^0_1\tilde\chi^0_3$ & only $\tilde\ell^+_R\tilde\ell^-_R$ & only $W^+W^-$ \\ \hline
    $\tilde\chi^0_1\tilde\chi^0_2\rightarrow\tilde\chi^0_1\tilde\chi^0_1\ell\ell$ ($\ell\neq\tau$) & 18343 & 615 & 51 & 855 \\
    $\tilde\chi^0_1\tilde\chi^0_3\rightarrow\tilde\chi^0_1\tilde\chi^0_1\ell\ell$ ($\ell\neq\tau$) & 290 & 20132 & 372 & 635 \\
    all SUSY background & 1153  & 3055 & 5626   & 951 \\
    all   SM background &   87  &  256 &   44   &  81  \\ \bottomrule
  \end{tabular}
\end{center}
\renewcommand{\arraystretch}{1.0}
\caption{Number of preselected events, 
  that fulfill the requirements of the kinematic selection procedure, for ${\mathcal{L}} = 500 \fb^{-1}$.}\label{tab:reconstruction_results}
\end{table}

The CP asymmetry can now be calculated from Eq.~\eqref{eq:asyth} to be
${\mathcal{A}}({\mathcal T} )_{\tilde\chi^0_1\tilde\chi^0_2} = -11.3\%
\pm 0.7\%$ and ${\mathcal{A}}({\mathcal T}
)_{\tilde\chi^0_1\tilde\chi^0_3} = +10.9\% \pm 0.7\%$.  The absolute
values are slightly higher than the ones calculated in the benchmark
scenario, since the asymmetry depends non-trivially on the cut values.
This has been studied in~\cite{CP_Paper} and can be taken into account
in a parameter fit.

\subsection{Fit of the parameters in the neutralino sector}

In the final step of the analysis, the accuracy to determine the
parameters in the neutralino sector of the MSSM is estimated. These
are the six free parameters of the neutralino mass matrix $|M_1|$,
$M_2$, $|\mu|$, $\tan\beta$, $\phi_1$ and $\phi_\mu$. As input for the
fit a number of CP-even observables is used together with the measured
asymmetries (see Ref.~\cite{CP_Paper} for details): $\mneu{1} = 117.3
\pm 0.2 \gev$, $\mneu{2} = 168.5 \pm 0.5 \gev$, $\mneu{3} = 180.8 \pm
0.5 \gev$, $\sigma(\neu{1}\neu{2}) \times {\rm
  BR}(\neu{2}\to\tilde{\ell}_R\ell) = 130.9 \pm 1.4 \fb$,
$\sigma(\neu{1}\neu{3}) \times {\rm BR}(\neu{3}\to\tilde{\ell}_R\ell)
= 155.7 \pm 1.6 \fb$, $\sigma(\neu{2}\neu{2}) \times {\rm
  BR}(\neu{2}\to\tilde{\ell}_R\ell)^2 = 4.8 \pm 0.3 \fb$,
$\sigma(\neu{3}\neu{3}) \times {\rm
  BR}(\neu{3}\to\tilde{\ell}_R\ell)^2 = 26.3 \pm 0.7 \fb$ and
$\sigma(\neu{2}\neu{3}) \times {\rm BR}(\neu{2}\to\tilde{\ell}_R\ell)
\times {\rm BR}(\neu{3}\to\tilde{\ell}_R\ell) = 28.9 \pm 0.7\fb$. The
fitted values of the parameters of the neutralino mass matrix are
listed in Tab.~\ref{tab:fitresults}.
\begin{table}%[htbp]
\begin{center}
%\vspace{0.5cm}
\renewcommand{\arraystretch}{1.2}
  \begin{tabular}{cccccc} \toprule
   $|M_1|$ & $M_2$ & $|\mu|$ & $\tan\beta$ & $\phi_1$ & $\phi_\mu$  \\
   $150.0 \pm 0.7 \gev$ & $300 \pm 5 \gev$ & $165.0 \pm 0.3 \gev$ & $10.0 \pm 1.6$ & $0.63 \pm 0.05$ & $0.0 \pm 0.2$   \\ \bottomrule
  \end{tabular}
\end{center}
\renewcommand{\arraystretch}{1.0}
\caption{Results of the parameter fit.}\label{tab:fitresults}
\end{table}
It is remarkable that the moduli of the phases $\phi_1$, $\phi_\mu$
can also be determined with high precision, using the CP-even
observables alone. However, only an inclusion of CP-odd asymmetries in
the fit allows to resolve the sign ambiguities of the phases.  Without
the CP-odd asymmetries in the fit there is a twofold ambiguity,
$\phi_1 = \pm 0.6$, and even fourfold if $\phi_\mu \neq 0$. Thus, the
triple product asymmetries are not only a direct test of CP violation,
but are also essential to determine the correct values of the phases.

\section{Summary and conclusions}

The first full detector simulation study to measure SUSY CP phases at
the ILC has been presented. Triple products of the final state lepton
momenta in neutralino decays have been used as CP-odd observables.
Realistic collider conditions have been simulated and all relevant SM
backgrounds have been taken into account. A detailed cut flow analysis
has been performed, including the development of a kinematic selection
procedure that was used to obtain a very clean signal sample and to
distinguish events from different neutralino decays. In the chosen
benchmark scenario the asymmetry could be measured with a relative
precision of 10\% with 500\,fb$^{-1}$ of data. Finally, the parameters
of the neutralino mixing matrix have been fitted to CP-even and CP-odd
observables and the complex phases could be determined with a
precision of about 10\%.

\section*{Acknowledgments}

We would like to thank Steve Aplin, Mikael Berggren, Jan Engels, Frank
Gaede, Nina Herder, Jenny List, and Mark Thomson for very useful
discussions and help with the detector simulations.

% ****************************************************************************
% BIBLIOGRAPHY AREA
% ****************************************************************************

\begin{footnotesize}

% ----------------------------------------------------------------------------

\end{footnotesize}

% ****************************************************************************
% END OF BIBLIOGRAPHY AREA
% ****************************************************************************

\end{document}